%% file: main.tex
\def\year{2022}\relax
\documentclass[letterpaper]{article} 
\usepackage{aaai22}  
\usepackage{times}  
\usepackage{helvet}  
\usepackage{courier}  
\usepackage[hyphens]{url}  
\usepackage{graphicx} 
\usepackage{natbib}  
\usepackage{caption} 
\DeclareCaptionStyle{ruled}{labelfont=normalfont,labelsep=colon,strut=off} 
\usepackage{algorithm}
\usepackage{algorithmic}
\usepackage{amsfonts}
\usepackage{mathtools}
\usepackage{amsmath}
\usepackage{bm}
\usepackage{dsfont}
\DeclareMathOperator{\ExpDis}{\mathrm{Exp}\,}
\usepackage{cleveref}
\usepackage{tikz}
\usetikzlibrary{shapes.misc}
\tikzset{cross/.style={cross out, draw, 
         minimum size=2*(#1-\pgflinewidth), 
         inner sep=0pt, outer sep=0pt}}

\usepackage{amssymb}
\usepackage{booktabs} 
\usepackage{newfloat}
\usepackage{listings}
\floatstyle{ruled}
\newfloat{listing}{tb}{lst}{}
\floatname{listing}{Listing}
\title{Collaborative Optimization of the Age of Information under Partial Observability}
\author{
    Anam Tahir, Kai Cui, Bastian Alt, Amr Rizk, Heinz Koeppl
}
\affiliations{

    anamtahir.sos.tu@gmail.com

}

\begin{document}

\maketitle

\begin{abstract}
The significance of the freshness of sensor and control data at the receiver side, often referred to as Age of Information (AoI), is fundamentally constrained by contention for limited network resources.
Evidently, network congestion is detrimental for AoI, where this congestion is partly self-induced by the sensor transmission process in addition to the contention from other transmitting sensors.
In this work, we devise a decentralized AoI-minimizing transmission policy for a number of sensor agents sharing capacity-limited, non-FIFO duplex channels that introduce random delays in communication with a common receiver. 
By implementing the same policy, however with no explicit inter-agent communication, the agents minimize the expected AoI in this partially observable system.
We cater to the partial observability due to random channel delays by designing a bootstrap particle filter that independently maintains a belief over the AoI of each agent.
We also leverage mean-field control approximations and reinforcement learning to derive scalable and optimal solutions for minimizing the expected AoI collaboratively.
\end{abstract}
\input{sections}

\bibliography{references}
\end{document}

%% file: sections.tex
\section{Introduction}
The Age of Information (AoI) is critical for decision-making and control in cyber-physical systems. AoI is a measure that quantifies the freshness of information of a sender, e.g., a sensor, calculated using the time elapsed since the last update message was received at the receiver. It is an important metric in real-time applications such as UAV-assisted communications, Internet-of-Things, wireless sensor networks, information processing systems, and cooperative, connected automated mobility~\cite{yates2021age, talak2017minimizing, hu2020aoi, wang2023survey, wang2023cooperative, baiocchi2021age, abd2019role}. 

In an AoI-based system, multiple sensors use the same channel, which has limited resources, to send their messages, resulting in the need for congestion control and scheduling algorithms to regulate the network traffic while minimizing the AoI. 
Various scheduling algorithms have been presented for this purpose, see \cite{muhammad2021minimizing, mankar2021throughput, yates2018age, hsu2017age, shao2021partially} and references therein. 
Not only is the channel of limited capacity, but it usually induces random delays on the messages that are transmitted such that these may arrive out of order.
This results in the partial observability of the state of the system since the sensors/agents do not possess \emph{instantaneous} information about their updated AoI at the receiver.
In this work, we present a bootstrap particle filter \cite{chopin2020introduction} which uses these delayed out-of-order messages to maintain a belief over the AoI of the agent.
Partial observability is also established as in such a  decentralized system, individual agents are not aware of the state of all other agents or the state of the entire system.

Partial observability in AoI-based systems has been considered in recent works, where the system is modelled as a (partially observable)-Markov decision process (PO-MDP) and then solved using methods from (deep) reinforcement learning (RL).
Authors in \cite{chen2020age} propose a proactive deep reinforcement learning algorithm to optimize the performance of a vehicle-to-vehicle network for AoI-aware radio resource management in vehicular networks, where the partial observability comes from agents working in a decentralized manner. 
In \cite{leng2019age} authors use the partially observable Markov game framework to model a decentralized wireless communication network and apply deep Q-learning to find the scheduling and power control policy for minimizing the average AoI, similarly~\cite{leng2021actor, ceran2021reinforcement, abd2019deep}.
Decentralized POMDP (Dec-POMDP) is a state-of-the-art framework for modeling these coordination problems \cite{oliehoek2016concise} by explicitly considering the uncertainty and partial observability of the environment.
However, this framework can be computationally expensive, especially as the number of agents and/or states increase \cite{bernstein2002complexity}, which is why not many AoI-based systems have used it so far, see \cite{song2022fast}.
A recently popular scalable method for multi-agent systems is mean-field approximations, which has also been used to learn policies for AoI-based systems~\cite{zhou2023age, wang2022dynamic, baiocchi2021age}.
Note that mean-field control, which is the collaborative mean-field formulation, has successfully been used to model multi-agent systems in the past \cite{carmona2019model, gu2021mean}, but is still new for AoI-based optimization systems.

In this work, we develop a collaborative algorithm adapted from a newly introduced decentralized partially observable mean-field control framework presented in \cite{cui2023learning} to model our AoI-based multi-agent system as a single agent MDP. 
Our main contributions through this work are: (i) We model a  system where the agents obtain delayed acknowledgements about their AoI leading to partial observability. (ii) We design a particle filter to cater to this uncertainty by maintaining a belief over the true AoI, i.e. the AoI as it is at the receiver. (iii) We adapt a partially observable mean-field control framework to learn scalable and collaborative policies for the decentralized system.

\begin{figure}[t]
\centering
\includegraphics[scale=0.35]{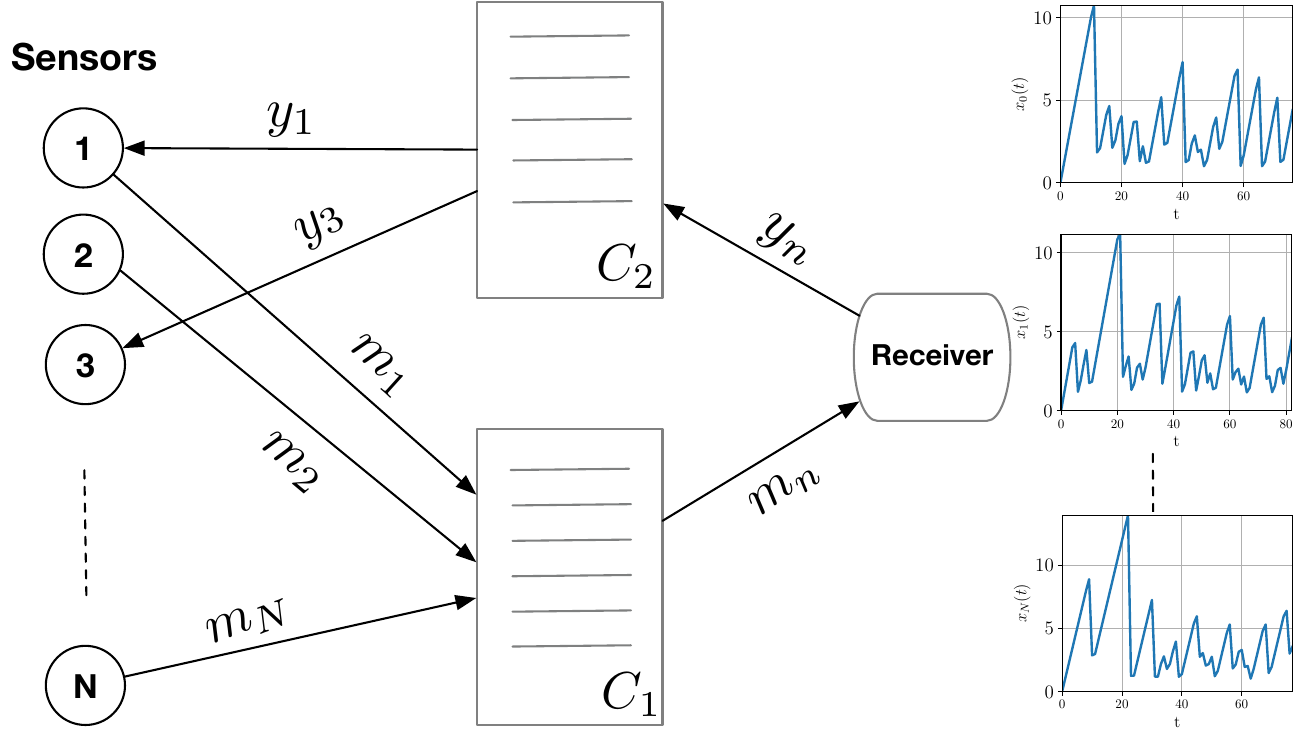}
\caption{Collaborative AoI system model consisting of $N$ agents, where each sensor (agent), $n \in \mathcal N$, sends its fresh message, $m_n$ through the finite capacity forward channel $C_1$ and receives the acknowledgements $y_n$ over the  finite capacity backward channel $C_2$. All agents share the goal of minimizing the own AoI at the receiver. The acknowledgements $y$ contain the correspondingly updated AoI $z$ and the update timestamp $\eta$.}
\label{fig:ma_system_model}
\end{figure}

\section{System Model}
We consider a multi-agent system having $N$ sensors, i.e. agents, which all aim to send their status updates as messages, $m_n$ for $m \in \mathbb N$ and $n \in \mathcal N$ with $\mathcal N = \{1, \ldots, N\}$, to a single receiver, through a finite capacity channel $C_1$.
The AoI of an agent $n$, $x_n(t) \in \mathbb R_{\geq 0}$, describes how old is the last received fresh update message at the receiver from that agent.
Our system model is illustrated in Fig.\ref{fig:ma_system_model}.
Next, we illustrate the dynamics of all components of this figure.

\subsection{Receiver Dynamics}
Upon receiving a fresh message from an agent $n\in \mathcal N$, the receiver updates the age of information (AoI) of agent $n$, independently of the other agents \cite{yates2021age}.
A fresh message is defined as a new arriving message $m_n$ with an index $m$ larger than all the previously received message indices of agent $n$.
The age of information process of each agent $n$, is given as $x_n \coloneqq \{x_{n}(t) \mid t \in \mathbb R_{\geq 0}\}$ as
\begin{equation}
    x_{n}(t) = t - \max_{m \in \mathbb N}(\tau_{n,m} \mid \tau_{n,m}' \leq t)
    \label{eq:aoi}
\end{equation}
where $t \in \mathbb R_{\geq 0}$ denotes the time  and $\tau_{n,m}$ is the sending time of the message with index $m$ of agent $n$ into the channel, with $m \in \mathbb N$.
We denote the reception time of the message $m_n$ at the receiver by $\tau_{n,m}'$.
The message $m$ is transmitted through channel $C_1$, which incurs a random delay that we assume to be sampled from an Exponential distribution $ \ExpDis (\lambda_1)$ with parameter $\lambda_1$, i.e., the one way delay $d_{n,m}^f \sim  \ExpDis(\lambda_1)$. 
The superscript indicates the channel index in the direction from the sensor agents to the receiver, so $f$ denotes the delay in \textit{forward} direction. 
This leads to the following reception time 

\begin{equation}
    \tau_{n,m}' = \tau_{n,m} + d_{n,m}^f.
    \label{eq:delay_model}
\end{equation}

Here we assume that the channel does not enqueue the packets but the agent is able to directly transmit the message upon creation. 
Note that, the reception times are not necessarily ordered, i.e., it might well be the case that $\tau'_{n,m} \nleq \tau'_{n,m'}$, for $m'>m$. 
The channel delay causes out-of-order delivery of messages to the receiver from each agent~\cite{rizk_palm}. 
The receiver, however, discards the outdated messages and only uses the newest ones to update the AoI using \eqref{eq:aoi}.
A message from an agent $n$ is fresh if its sending time $\tau_{n,m}$ was the latest as compared to all the messages received so far by the receiver from this agent.

In the following, we assume a measurement model at the receiver of the process for agent $n$, $x_n(t)$ at time points $\eta_n \coloneqq \{\eta_{n,j}\}_{j \in \mathbb N}$ when a fresh message arrives, i.e.,

\begin{equation}
    \eta_{n,j}=\left\{\tau_{n,j}' \middle| j=\underset{m \in \mathbb N}{\arg\max}(\tau_{n,m}\mid { \tau_{n,m}' \leq t} ) \land \forall t \in \mathbb R_{\geq 0}\right\}
    \label{eq:rcvng_time_of_non_obs_msg_at_rcvr}
\end{equation}

where $j \in \mathbb N$. We assume noise-free measurements $\{z_{n,j}\}_{j \in \mathbb N}$ of the process $x_n$: $z_{n,j}=x_n(\eta_{n,j})$.
The receiver immediately transmits an acknowledgement of each received fresh message back to the sender, containing the message reception time and the correspondingly updated AoI. 
These acknowledgements are received by the agent $n$ at time points $\eta'_{n,j}$, explained in detail in the partial observability section.
Note that we use the terms $ z_{n,j}$ and $x_n(\eta_{n,j})$ interchangeably due to the absence of noise.
A realization of this process can be found in Fig.~\ref{fig:sample} for agent $n$.

\input{example_realization_figure}

\subsection{{Channel Dynamics}}
The considered system consists of two independent finite capacity channels, a forward channel $C_1$ and a backward channel $C_2$.
The channel $C_1$ is used by all the agents to send their fresh messages to the receiver. 
In congruence with the delay model in \eqref{eq:delay_model} we consider a forward channel that consists of a number of paths $C$ that each can transmit one message at a time. 
We assume a fixed channel utilization (overbooking ratio), specifically, in terms of the ratio of the number of sensors to the number of paths, i.e.,  $C = \kappa N$, where $\kappa \le 1$. 
At any time point when all the paths in the channel $C_1$ are occupied, any new incoming message is dropped which is registered by the sender.
In the considered discrete-time agent action model the total number of messages admitted into the channel at every time step is at most equal to the total number of free paths.
We note that the system runs in continuous time, however, the agents observe the system and correspondingly take actions only at discrete time points.
The admitted messages are chosen uniformly at random out of all incoming messages.
Note that any agent can have more than one message on the channel at any time.
Each path in the channel delivers its message to the receiver after a delay, $d_{n,m}^f \sim \ExpDis(\lambda_1)$. At any time $t$, the state of a path in the channel is $s_{c}(t) \in \mathcal S = \{0,1\}$, for $c \in \{1,\ldots,C\}$, where $s_{c}(t) = 0$ indicates that the path is free and $s_{c}(t) = 1$ indicates an occupied path.  
The channel load, i.e., the ratio of occupied paths to $C$ at time $t$, is: 
\begin{align}
    \nu_1(t) = \frac{1}{C} \sum_{c=1}^{C} \mathds 1_{s_{c}(t)}.
    \label{eq:occupied_paths}
\end{align}
Channel $C_2$ is a designated channel used by the receiver to only send back acknowledgements to the sensor agents, hence the capacity $C$ of channel $C_2$ is set equal to that of channel $C_1$.
Note that the receiver only sends acknowledgements of the messages it uses to update the AoI and not for the outdated messages.
The channel delay model is $\ExpDis$ with parameter $\lambda_2$.

\subsection{Agent Dynamics}
An agent $n$ generates and immediately sends a fresh message to the receiver according to some policy that defines the inter-transmission times. 
Upon receiving a message, if it is not outdated due to out-of-order arrivals, the receiver updates the AoI for that agent at time points $\eta_n$, explained before.
The agents do not obtain individual information on other agents' states or policy. 
The only information an agent can access from the environment is the current channel load, i.e., how occupied is the channel in terms of $\nu_1(t)$ from \eqref{eq:occupied_paths}. 

Given the assumed delay distributions, the full state of an agent $n$ at time $t$ is defined as $X_{n}(t) = (x_{n}(t), M_{n}^1(t), M_{n}^2(t))$, where $x_{n}(t)$ is the current AoI for the agent $n$ at the receiver, $M_{n}^1(t)$ is the number of messages in channel $C_1$ from the agent $n$ and $M_{n}^2(t)$ is the number of messages (acknowledgements) in channel $C_2$ from the receiver to agent $n$. 
The agent $n$ receives observations, $y_{n}$, which essentially are the acknowledgements sent by the receiver. These acknowledgements, $y_{n}$, contain (i) the reception time $\eta_{n,j}$ of the fresh message for that agent, and (ii) the corresponding value of the updated AoI $x_{n}(\eta_{n,j})$ at receiver.

In this work, we assume that the agents do not have this full information. 
We assume that an agent only has the count of the number of its unacknowledged messages, $u_{n}(t) = M_{n}^1(t) + M_{n}^2(t)$, and it can keep a belief over $x_{n,t}$ using the received observations $y_{n}$ and the parameters of the channel delay model.
As we will show later, we assume that the agent does not observe the continuous time processes $u,x$ but rather samples its observations at discrete times and also takes actions on this discrete time scale. 
The actions, which will be illustrated in detail in the system design section, are mainly whether to generate and send a fresh message in this discrete time step or not. 
Finally, we assume that the agent $n$ can only obtain the state of the channel $C_1$ when it sends to it, irrespective of whether its message is dropped or not.
This leads to the considered system being partially observable as discussed in the next section.

\paragraph{\textbf{Reward function}}
The agents aim to learn a policy, $\pi$ that decides whether to generate and transmit fresh messages, in order to maximize a reward. 
Since we have limited capacity channels being shared by all agents, we assume that the agents are willing to implicitly cooperate in order to maximize the global reward, which is to minimize the expected AoI over all agents.
Hence, we assume the following reward function:
\begin{align}\label{eq:reward}
    R(t) = - \frac{1}{N} \sum_{n \in \mathcal N} x_{n}(t)
\end{align}
where $\mathcal N = \{1, \ldots, N\}$.
For instance, one can simply penalize message drops in the reward function. 
Since the agents are working in an implicit cooperative manner to maximize this reward, we assume that this global reward is distributed to all agents.

\section{{AoI} Minimization under Partial Observability}

Next, we first illustrate the effect of the random channel delay on the observation model of the agents.
We then propose a bootstrap particles filter for the agents in order to cater for the delayed observations.
Lastly, we introduce the partially observable mean-field framework that is used to obtain decentralized policies for the agents to minimize their AoI.

\subsection{{Agents' Partial Observability}} 
Recall, that the measurements from the receiver $\{z_{n,j}\}_{j \in \mathbb N}$ are not immediately observed by the corresponding agent $n$. 
The receiver sends the acknowledgement over the channel $C_2$, where each message is subjected to a random delay.
The observations $y_{n}$, i.e., the delayed acknowledgement messages, are received at time points
\begin{equation}
    \eta'_{n,j} = \eta_{n,j} + d_{n,j}^r,
    \label{eq:rcvng_time_of_msg_at_agent}
\end{equation}
with random i.i.d. delay times $ d_{n,j}^r \sim  \ExpDis(\lambda_2)$, with parameter $\lambda_2$.
The superscript $r$ denotes the delay in \textit{reverse} direction. 
Note that due to this channel delay, the acknowledgements received by the agents can be out-of-order. 
However, we make use of all the received observations to update our particle filter.

The ordered sending times of the receiver for acknowledgements of agent $n \in \mathcal N$, denoted  $\rho_n=\{\rho_{n,j}\}_{j \in \mathbb N}$ are then given as the order statistic of the sequence $\eta_n = \{\eta_{n,j}\}_{j \in \mathbb N}$, i.e.,
\begin{equation}
    \{\rho_{n,k}\} =\{\eta_{n,(k)}\} \quad \forall k \in \mathbb N,
    \label{eq:ordered_send_times_at_rcvr}
\end{equation}
where $(\cdot)_{n,(k)}$ denotes the $k$th order statistic, or the $k$th smallest value, of the set $\eta_n$. Additionally, $l(k):\, \mathbb N \rightarrow \mathbb N$ denotes the corresponding index of this value in the set $\eta_n$ \cite{david2004order}.

Then the sequence of observations $y_n \coloneqq \{y_{n,j}\}_{j \in \mathbb N}$ is given as
\begin{equation}
    \{y_{n,j}\}_{j \in \mathbb N}=\{z_{n,l(k)}\}_{k \in \mathbb N}.
    \label{eq:seq_of_obs}
\end{equation}
Similarly, the ordered sequence of sending time points is then determined at the agent as:
\begin{equation}
    \{\rho_{n,j}\}_{j \in \mathbb N}=\{\eta_{n,l(k)}\}_{k \in \mathbb N}.
    \label{eq:ordered_seq_of_obs}
\end{equation}

We use the information contained in the acknowledgements to feed  the particle filter \cite{chopin2020introduction} of each agent to independently update its belief over its own AoI. 
The agent then also estimates the delay incurred by  channel $C_2$ using \eqref{eq:rcvng_time_of_msg_at_agent} and learns the delay distribution parameter. 
We assume that the agent correctly knows the delay distribution class. 
Finally, note that the agent does not maintain a belief over $\nu_1(t)$ although it only obtains the state of channel $C_1$ when it sends to it.
Calculating this belief is non-trivial and intractable, since each agent then needs to know the policy or the last timestep action of other agents.
Further, we give a concise explanation of each of the time variables used in this paper, both at the receiver and agent end. 

For the agent, we have the following:
\begin{itemize}
    \item $\tau_{n,m}$: time at which agent $n$ successfully send message number $m$ to the receiver.
    \item $\eta'_n \coloneqq \{\eta'_{n,j}\}_{j \in \mathbb N}$: times when the acknowledgements are received by the agent $n$. Due to channel delay, it can be that: $\eta'_{n,j} \nleq \eta'_{n,j'}$, for $j'>j$ and $\{j,j'\} \in \mathbb N$.
    \item $\rho_n=\{\rho_{n,j}\}_{j \in \mathbb N}$: the ordered sequence of the times, $\eta_{n,j}$, the acknowledgements $j$ were sent by the receiver to agent $n$.
    
\end{itemize}
And at the receiver, we have:
\begin{itemize}
    \item $\tau'_{n,m}$: time at which message $m$ of agent $n$ is received by the receiver. Due to channel delay, it can be that: $\tau'_{n,m} \nleq \tau'_{n,m'}$, for $m'>m$ and $\{m,m'\} \in \mathbb N$.
    \item $\eta_n \coloneqq \{\eta_{n,j}\}_{j \in \mathbb N}$: times when a fresh message arrives at the receiver from an agent $n$ and is used to update the AoI of the agent $n$ at the receiver.
\end{itemize}

\input{particle_filter_figure}

\subsection{{Particle Filter}}
As the agents receive delayed acknowledgements about their individual AoI,  the system is partially observable.
In order to make informed decisions, the agents maintain a belief over their AoI state using a particle filter~\cite{sarkka2013bayesian, chopin2020introduction}.
{For ease of notation, we have removed the subscript $n$ from the following description that is without loss of generality shown for one agent.}

Let $I$ denote the number of observations obtained by the agent until time point $t$, i.e., the agents' observation time points $\eta'_1<\eta'_2<\dots<\eta'_I<t$ and the corresponding receivers' ordered sending times $\rho_1<\rho_2<\dots<\rho_I<t$ from \eqref{eq:ordered_seq_of_obs}. 
We compute the posterior distribution of the latent path of the AoI process  $x_{[0,t]}$ given the observations $y_{1:I} \coloneqq \{y_i\}_{i=1}^I$, their time points $\eta'_{1:I} \coloneqq \{\eta'_i\}_{i=1}^I$ and the sorted sending times $\rho_{1:I} \coloneqq \{\rho_i\}_{i=1}^I$.
We denote a path of $x$ between two time points $t<t'$ as $x_{[t,t']}\coloneqq\{x(s)\mid t \leq s \leq t'\}.$ 
This posterior inference $p(x_{[0,t]} \mid y_{1:I}, \rho_{1:I}, \eta'_{1:I})$ is hard to compute, hence, we use a particle filter to represent this distribution in terms of a set of $P$ particles,  $x^p$ and their associated likelihood weights, $w^p$, for $p = \{1, \ldots, P\}$.
The likelihood, for our noise-free model, is given by a discrete distribution as: 
\begin{equation}
\begin{aligned}
    p( y_{K} \mid x(\rho_K))&= \delta( y_K - x(\rho_K))
\end{aligned}
\label{eq:likelihood}
\end{equation}
where $K$ is the number of observations in the interval $[0,\eta'_I]$ and $x(\rho_K)=z_{(l(k))}$.

We use the bootstrap particle filter in which the importance distribution is assumed to be the same as the prior distribution, so the weights only depend on the likelihood which is given in~\eqref{eq:likelihood}, see \cite{sarkka2013bayesian, chopin2020introduction} for more details. 
This is hard to achieve given the Dirac form of the likelihood in \eqref{eq:likelihood} when the filter updates the particles, especially if the state is continuous. 
Hence, the particles are reset to the last received observation, $y_K$, and their weights remain equally likely after every update.

We explain the designed particle filter using the example, depicted in Fig.~\ref{fig:belief_update}.
Given the acknowledgement/observation $y_k$ for $k\in \mathbb{N}$, containing the pair $\{z_{k}, \rho_{k}\}$, the value of $x(\rho_k)$ associated with the measurement time point $\eta_j$ is now known to the agent and does not need to be estimated. 
Once an observation is received, irrespective of the sequence order, all the particles of the agent are updated by going back in time to $\rho_k$ and updating the value of all particles to $z_{(l(k))}$.

In Fig. \ref{fig:belief_update} we show two such particles $x_{1}(t)$ and $x_{2}(t)$. 
These start at a time $t=0$ using the initial belief, so $x_1(0)=x_2(0)=0$. 
The particles are \emph{simulated} at a rate of $1$ (as the AoI advances) until a time point where a sent message $m_j$ may have been received by the receiver. 
The solid line is the simulation of the particle filter without observations and knowing the delay model distributions.
The particles keep propagating in this manner until an observation $y_j$ is received at the time $\eta'_j$. 
Upon receiving $y_j$, each particle in the set $P$ is back-propagated to $z_{(l(k))}$ at time $\rho_j$, using \eqref{eq:seq_of_obs} and \eqref{eq:ordered_seq_of_obs}.
The dashed lines are simulations based on a back propagated observation that was received. 
This simulation from this back-propagated point on is too based on the known delay distributions until the time $t$.

\begin{figure}
\centering
\includegraphics[scale=0.19]{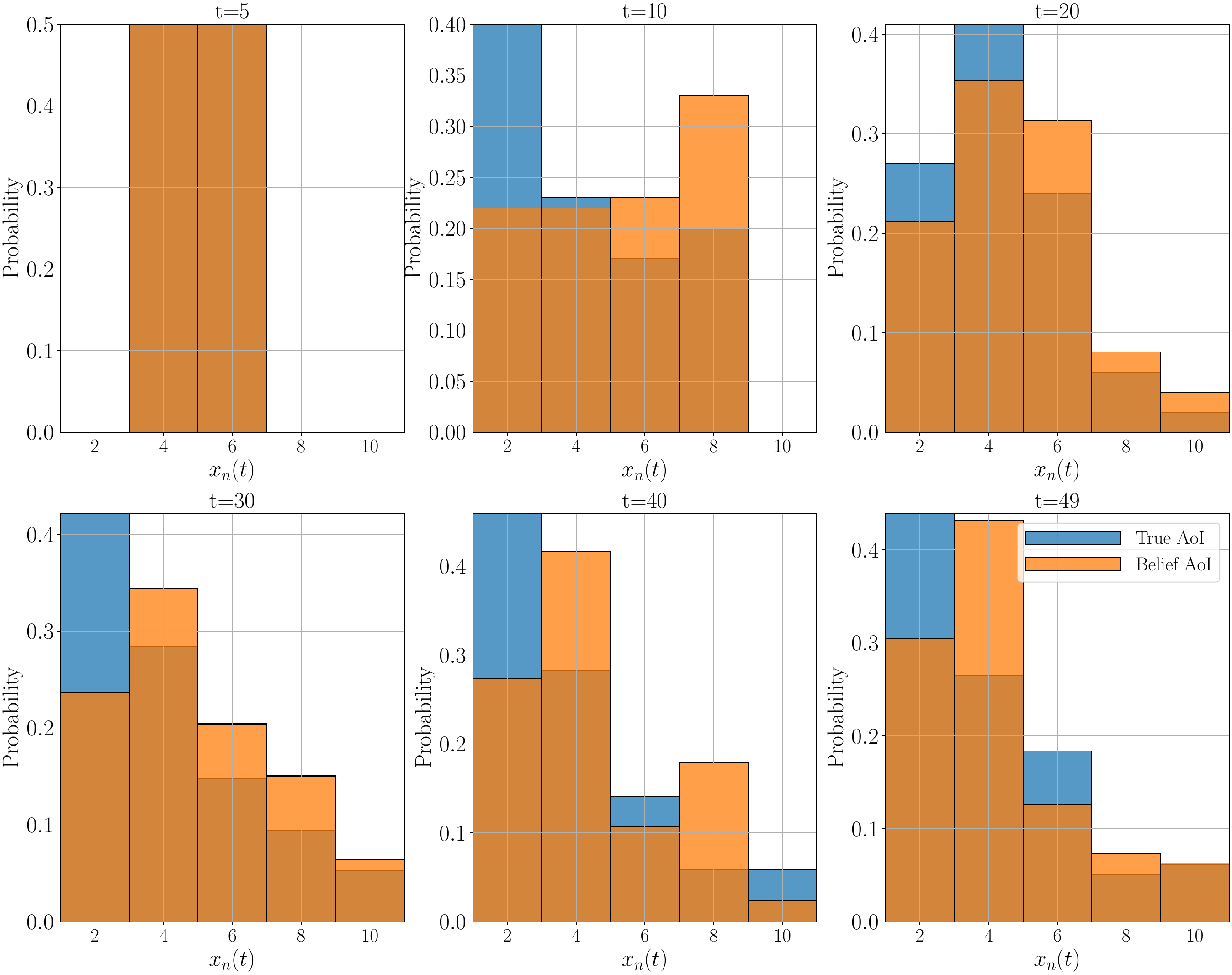}
\caption{Subplots show empirical histograms for the true AoI (blue) and the expected belief AoI (orange) for all agents, with $N=100$ with evaluation done at times $t=\{5,10,20,30,40,49\}$ seconds.
Observe that the belief AoI follows the pattern of the true AoI.
}
\label{fig:pf_hist}
\end{figure}

Using the state,  $x_{p}(t)$ where $p = \{1,2,3, \ldots, P\}$, of all its particles, the agent computes the mean, $\mu_P(t)=\frac{1}{P} \sum_{p=1}^P x_{p}(t)$, and standard deviation, $\sigma_P(t) = \sqrt{ \frac{1}{P} \sum_{p=1}^P (x_{p}(t) - \mu_p(t))^2}$, at any time $t$.
In Fig. \ref{fig:pf_hist} the empirical distribution of the AoI of all $N=100$ agents is given in the form of histograms. 
The blue histogram is for the true AoI at the receiver and the yellow histogram is for the mean belief AoI, $\mu_P(t)$, maintained at each agent using our proposed particle filter. 
The initial true state and particles are all set to $0$ and then the performance evaluation is done at different times, $t=\{5,10,20,30,40,49\}$. 
Using our proposed filter the agents are able to maintain a belief over the AoI which is close to the true AoI at the receiver. 
Recall that this true AoI is unknown to the agent except when a delayed acknowledgement is received.

\begin{figure}
\centering
    \includegraphics[scale=0.4]{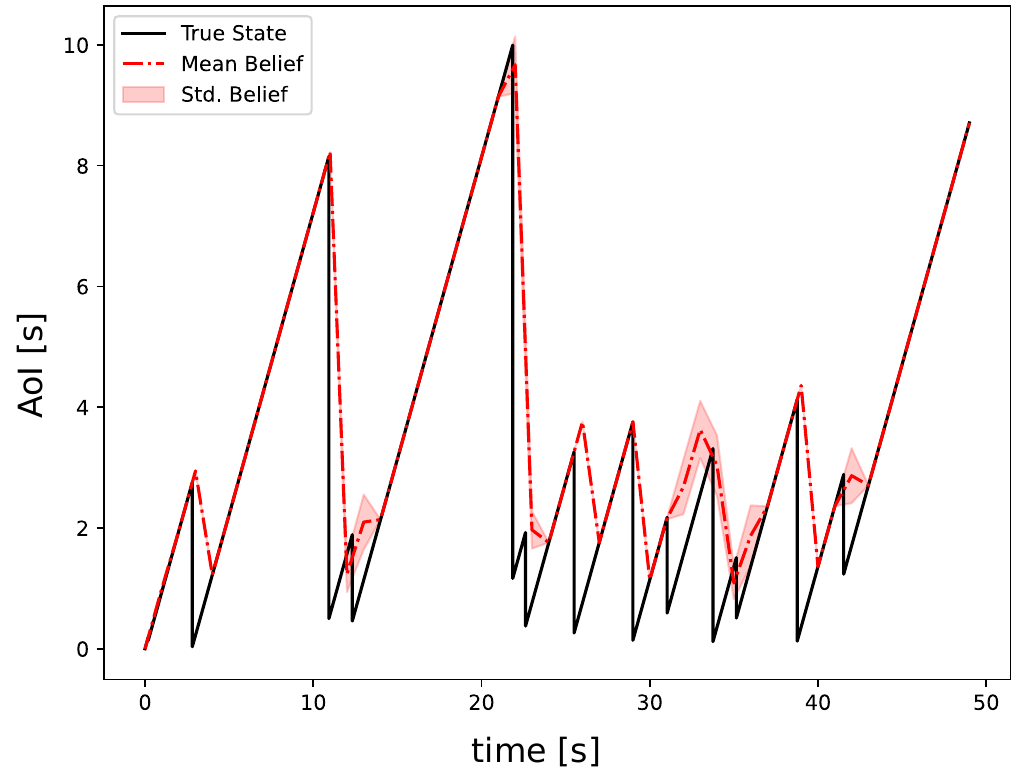}
\caption{Realization of our bootstrap particle filter estimates over a time period of $50$ seconds (x-axis) for one agent. The agent mean belief state (dotted red line) is close to its true AoI state at all times (black). Recall that acknowledgements are received after random delays.}
\label{fig:particle_filter_perf}
\end{figure}

Fig. \ref{fig:particle_filter_perf} shows the performance of our particle filter for one agent over a time period of $50$ seconds. It can be seen that the agent belief of their AoI (red) is close to the true AoI at the receiver (black). 
This indicates that we can use the belief state to learn a good control policy for the agent actions.
Since the channel state depends on the joint actions of all agents, it is not possible for each agent to use a similar approach to keep a belief over the channel state independently. So to keep things tractable we will instead use only the last received individual channel state information by each agent to learn a decentralized policy.

\subsection{{Partially Observable Mean-Field}}
Mean-field representations are a successful way of learning scalable solutions in multi-agent systems, both in cooperative \cite{gu2021mean} and competitive settings \cite{yardim2023policy}. 
Our system model fits the former since we aim to reduce the expected AoI of the entire system. 
Note that learning a decentralized mean-field policy in partially observable systems still remains relatively uncharted, see \cite{cui2023learning} its references.

We adapt the work of \cite{cui2023learning} which presents a decentralized partially observable mean-field control (Dec-POMFC) model and assumes permutation-invariant agents to reduce it to a single agent Markov-decision Process (MDP). 
A scalable policy for this MDP is then learned using the state-of-the-art single-agent RL policy gradient method known as Proximal Policy Optimization (PPO). 
In essence, the approach assumes the usage of "upper-level" policy, $\tilde \pi$, during training, which assigns for any agent the probability with which it should send updates under its current belief of AoI. Note that the policy $\tilde \pi$ is for the mean-field system.

As mentioned in our agent dynamics section, the full state of the agent $n$ at time $t$ can be defined as $X_{n}(t) = (x_{n}(t), M_{n}^1(t), M_{n}^2(t))$ and the observations are denoted by $y_{n}$. 
For the Dec-POMFC representation, a joint distribution of all agents, $\mu(t) = \frac{1}{N}\sum_i \delta_{X_{n}(t)}$, over the three components of the state $X_{n}(t)$ is considered.
The number of total messages in channel $C_1$, denoted by $J_{C_1}(t)$, is instantaneously observed by an agent only when it sends a message to channel $C_1$, but can be calculated from the full state of all agents at any time $t$, as $J_{C_1}(t) = \sum_{i=1}^N M_{i}^1(t)$.
From this, the ratio of occupied paths in \eqref{eq:occupied_paths} is obtained as $\nu_1(t) = \frac{J_{C_1}(t)}{C}$, and the empirical distribution over channel fullness is calculated as $\nu = [\nu_0, \nu_1]$, where $\nu_0 = 1 - \nu_1$.

To learn the upper-level policy, $\tilde \pi$, for the Dec-POMFC model, in contrast to \cite{cui2023learning}, we make use of the mean and standard deviation of all three components of the state, $X_{n}(t)$ for all $n \in \mathcal N$ agents and $\nu(t)$ as a lower-dimensional and efficient input representation of the mean-field. Note that the mean and standard deviation are just a specific function of the true distribution, $\mu(t)$, and can be a good approximation to the otherwise computationally intractable $\mu(t)$. 
The learned upper-level policy, $\tilde \pi$, can then be used to extract the lower-level policy for each agent, $h_{n}(t) \sim \tilde \pi(\mu(t))$, i.e. the probability to send update messages given any current AoI, which decides on the agent actions $a_n(t) \sim h_{n}(t, x_{n}(t))$.
Recall that the reward is given in \eqref{eq:reward}. 
Note that although the upper-level policy $\tilde \pi$ accesses the full system information $\mu(t)$ during training, we can perform decentralized execution, i.e., without knowledge of $\mu(t)$, after training the policy \cite{cui2023learning}.

Now, for the case when agents don't have up-to-date information on their AoI at all times, the particle filter is used to maintain a belief over the AoI \cite{sarkka2013bayesian}. 
This belief is represented by each agent using a set of particles, which can then be used instead of the true AoI $x_{n}(t)$ as discussed.

\section{System Design}
Next, we give our system design in terms of the action space and observation space, the values of the parameters used, as well as, a description of the solver.
We first explain our training process for learning the policy and how it is then evaluated on the system when the $N$ is large.

\paragraph{\textbf{Action space}} The agents learn a policy, $\pi$, that dictates whether to generate and send a new message at the current timestep or not. 
So, the action space consists of, $a \in \mathcal A = \{0, 1\}$, where $a=1$ denotes a new message is generated and sent at this time step. 
The policy should tell the agent what action to take based on the state the agent is in, here, the state is the known or the belief AoI of each agent. Since the AoI is a continuous random variable, $x_{n}(t) \in \mathbb R_{\geq 0}$, the state space will be large leading to a long training time or a non-optimal learned policy.
To resolve this, we quantize the AoI $x_{n}(t)$ into finite $q$ levels, $\{[0,1), [1,2), [2,3), \ldots, [q-2,q-1), [q-1,\infty]$, which dictates then the size of our action space for each agent $n$ given as $\mathbf a_{n} = \{a^1_{n}, a^2_{n}, \ldots, a^{q}_{n}\}$, where $a^i_{n} \in \{0,1\}$ for $i=1,\ldots,q$, is an action for each quantized state level.
Note that any finite number of quantization levels can be used.

\paragraph{\textbf{Observation space}} We consider the following three models for learning the policy as they differ in terms of the information used by the agents for training.
\begin{itemize}
    \item \textbf{POMFC model}: This is the lower-dimensional representation of our partially observable mean-field representation Dec-POMFC model, discussed in system model section. To learn the POMFC policy
    the input observation is a vector containing the mean and standard deviation over the full state of all agents and channel state information, $\mathbf o=(\mu(x_{n}(t)), \mu(M_{n}^1(t)), \mu(M_{n}^2(t)),\\ \sigma(x_{n}(t)), \sigma(M_{n}^1(t)), \sigma(M_{n}^2(t)),    \nu_1(t))_{n=1, \ldots, N}$. Once an upper-level policy $\tilde \pi$ is learned it is used to extract the lower-level policy $h_n(t)$ for each agent from which the agent can choose its actions based on its individual AoI, $a_n(t) \sim h_n(x_n(t))$.
    \item \textbf{NA model}: This is the N-Agent scenario where the AoI of each agent is assumed known and directly used to learn the policy, leading to the input observation: $\mathbf o = (x_{n}(t))_{n=1,\ldots,N}$. Note that the input observation of this model is the same as that used by the lower-level policy $h_n(t)$ in our proposed POMFC model.
    \item \textbf{NA-Dec model}: This is the N-Agent scenario that uses the channel state information and the total unacknowledged messages of each agent to learn the policy. The input observation is: $\mathbf o = (x_{n}(t), u_{n}(t), \nu_1(t))_{n=1,\ldots,N}$, where $u_{n}(t) = M_{n}^1(t) + M_{n}^2(t)$ is the total unacknowledged messages of each agent.
\end{itemize}

Based on the channel-induced delay we consider the following two variations of observations
\begin{itemize}
    \item \textbf{TrueState} version: This is the scenario where we assume that the agent perfectly observes its true AoI, $x_n(t)$ at the receiver instantaneously without delay and also knows about the channel state, $\nu_1(t)$, at all times.
    \item \textbf{AvgBelief} version: This is the scenario where the channel-induced delay is taken into account and the agents only have access to delayed acknowledgements in the form of observations. Each agent then uses the particle filter, from the particle filter section, to maintain a belief over its true AoI which is used to calculate the average belief, $\mu_{n,P}(t)$, and the standard deviation, $\sigma_{n,P}(t)$, of each agent $n=1,\ldots,N$ and hence the entire system.
    Additionally, for the channel state, we  only use the last observation of each agent which they received upon the last attempt to use the channel.
\end{itemize}

For the NA-Dec model, we use as input observation all the particles directly, instead of their mean and standard deviation as in the AvgBelief case. The input observation for this NA-Dec-Particles version is then given as $\mathbf o = (x_{n,p}(t), u_{n}(t), \tilde \nu_{n,1}(t))_{n=1,\ldots,N, p=1,\ldots,P}$. 
We use this model to understand if using the lower-dimensional representation (mean and standard deviation) is enough to learn a well-performing policy.

\begin{figure}
\centering
\includegraphics[scale=0.8]{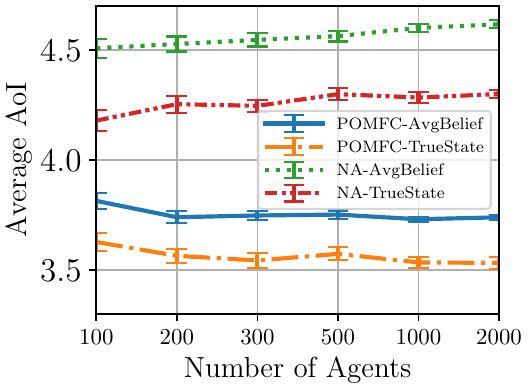}
\caption{Performance comparison of both POMFC policies to its finite system versions, NA. POMFC policies make use of additional, implicit information on the channel state to obtain better policies. }
\label{fig:pomfc_vs_na}
\vspace{-10pt}
\end{figure}


\begin{algorithm}
\caption{Application of POMFC policy in finite system}
\label{alg}
\textbf{Input}:  Trajectory of decision rules $h(t) \sim \tilde \pi(\mathbf o)$ for $t=0,\ldots,T_e$ \\
\textbf{Parameter}: Given in supplementary material
\begin{algorithmic}[1] 
\FOR {$n = 1, \ldots, N$}
\STATE Initialize states: $x_{n}(0)=M_{n}^1(0)=M_{n}^2(0)=0$.
\STATE Initialize $P$ particles: $x_{n,p}(0)=0$, for $p=\{1,\ldots,P\}$.
\ENDFOR
\FOR {$c = 1, \ldots, C_1$}
    \STATE Initialize channel states $s_{c}(0)=0 $.
\ENDFOR
\FOR  {$t = 0, 1, \ldots, T_\mathrm e$}
    \FOR  {$n = 1, \ldots, N$}
        \STATE Get agent state $x_{n}(t)$ 
        \STATE Compute quantized state index $q_{n}(t)$
        \STATE Get agent action $a_{n}(t) \sim h(t)(t, q_{n}(t))$.
    \ENDFOR
    \STATE Propagate the system using $\mathbf a_{n}(t)$ for timestep $\Delta t$
    \STATE Count number of dropped packets
    \STATE Calculate the average AoI for all agents.
    \FOR {$n = 1, \ldots, N$}
        \STATE Update the state $X_{n}(t) = (x_{n}(t), M_{n}^1(t), M_{n}^2(t))$
        \STATE Update set of particles using bootstrap filter
    \ENDFOR
\ENDFOR
\RETURN Average AoI and dropped packets.
\end{algorithmic}
\end{algorithm}

\paragraph{\textbf{Solver}} In order to learn a policy $\pi$ for each agent we use the state-of-the-art proximal policy algorithm~\cite{schulman2017proximal}, which has shown promising results for both discrete and continuous state and action spaces. The reward function~\eqref{eq:reward} focuses solely on minimizing the average AoI of the system.
To cater for partial observability of the channel and delayed observations of the acknowledgements, we use recurrent neural networks~\cite{yu2019review} with PPO. 
The experiments use the RLlib implementation~\cite{liang2018rllib} of PPO with the choice of design and hyperparameters given in Table ~\ref{table:hyperparameters}.
The design parameters used in all our experiments are given in Table ~\ref{table:parameters}.

\paragraph{\textbf{Evaluation}} Once a policy is learned it can then be used to extract the action $a_{n}(t)$ for an agent in a decentralized manner by using the AoI state (true or belief) $x_{n}(t)$ of that agent. 
This only requires the quantized index $q_{n}(t)$ which reflects in which quantization level, $\mathbf a_{n} = \{a^1_{n}, a^2_{n}, \ldots, a^{q}_{n}\}$, the AoI lies for the agent $n$.
The pseudocode for evaluating our POMFC in the finite system is given in Algorithm \ref{alg}, and the Python implementation of our system will be provided as open source.

\begin{table}[h]
    \centering
    \footnotesize
    \begin{tabular}{@{}ccc@{}}
    \toprule
    Symbol     & Name          & Value     \\ \midrule
    $\gamma$ &   Discount factor &  $0.99$\\
    $\lambda_\mathrm{RL}$ &   GAE lambda &  $1$\\
    $\beta_c$ &   KL coefficient & $0.2$ \\
    $\beta_t$ & KL target & $0.01$ \\
    $\epsilon$ &  Clip parameter & $0.3$ \\
    $l_{r}$ &   Learning rate & $0.00005$ \\
    $B_{b}$ &  Training batch size &  $4000 - 24000$ \\
    $B_{m}$ & SGD Mini batch size &  $128-4000$ \\
    $I_{m}$ & Number of SGD iterations & $5-8$ \\
    $f_l$ &  FCNet hidden layers & $[256, 64]$ \\
    $l$ &  LSTM cell size & $50$ \\
    $f_a$ & FCNet activation function & tanh \\
    \bottomrule
    \end{tabular}
    \caption{Hyperparameter configuration for PPO.}
    \label{table:hyperparameters}
\end{table}

\begin{table}
    \centering
    \footnotesize
    \begin{tabular}{@{}ccc@{}}
    \toprule
    Symbol     & Name          & Value     \\ \midrule
    $\Delta t$   &  Discrete time step   & $1$  \\
    $N$          &    Number of agents   & $100-2000$   \\
    $\kappa$    & Ratio of agents to paths &   $0.5$ \\
    $C$          &    Number of channel paths  & $50-1000$   \\
    $\lambda_1$ & $C_1$ channel delay parameter & $1.5$ \\
    $\lambda_2$ & $C_2$ channel delay parameter & $1.5$ \\
    $P$          &    Number of particles  & $100$   \\
    $S$          &    Monte Carlo simulations   & $100$   \\
    $\nu_1(0)$ & Ratio of occupied paths at $t=0$ & $0$  \\
    $D$          &    Drop penalty per message   & $1$   \\
    $T$          &    Training episode length   & $50$   \\
    $T_\mathrm{e}$          &    Evaluation episode length   & $50$ \\
    $q$ & Quantization levels & $16$\\
    $\omega$ & ConstantRate & $0.1,0.2,\ldots,1.0$\\
    $\alpha$ & Threshold & $1,2,\ldots,8$\\
    \bottomrule
    \end{tabular}
    \caption{System parameters used in the evaluations.}
    \label{table:parameters}
\end{table}

\begin{figure}
\centering
\includegraphics[scale=0.72]{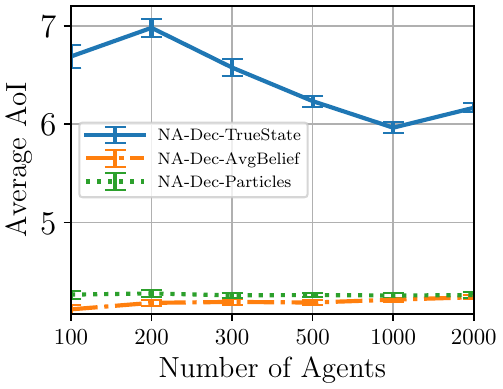}
\caption{Average AoI for different versions of NA-Dec shows that in comparison to only knowing the true AoI, using more information on the channel, as in NA-Dec-Particles and NA-Dec AvgBelief, leads to a better-learned policy.}
\label{fig:na_dec}
\end{figure}

\section{Evaluation Results}
In this section, we present the results of our experiments. 
As stated above we use PPO to learn a policy that maximizes the reward given in~\eqref{eq:reward}. 
This training method falls in the category of centralized training and decentralized execution (CTDE) \cite{kraemer2016multi}, with parameter sharing \cite{pham2018efficient}, in order to circumvent the non-stationarity issue common in the MARL setup.

For all the results we present below, the learned policy is evaluated on a range of agents: $N = \{10, 100, 200, 500, 1000, 2000\}$ together with error bars. 
For every, $N$ the evaluation comprises $100$ Monte Carlo simulations with corresponding confidence intervals. 
The POMFC policies were trained on $100$ agents, while the NA policies were trained using $10$ agents only. These were chosen because $100$ are agents are already enough to define a large system, even though POMFC framework is scalable, and a higher number of agents could have been used. However, for the NA framework, due to its lack of scalability higher number of agents was resulting in a much higher convergence time.
This reduction in the number of training agents was necessary since the training time for multi-agent systems increases drastically with the number of agents.
All the policies are compared based on the \emph{average AoI} in the system, i.e.,the average of the AoI of all $N$ agents, which reflects the reward function.

In addition to the above introduced learned policies we have also compared to the following \textbf{fixed policies}: (i)  \textbf{ConstantRate}: based on a fixed rate, $\omega$, a subset of agents is randomly chosen to send their messages at every timestep.  (ii) \textbf{AlwaysSend}: every agent is sending a fresh message at every timestep regardless of the AoI or channel state. This is bound to perform badly, especially in terms of message drops in the system. (iii)  \textbf{Threshold}: every agent sends a fresh message if their true AoI exceeds certain threshold, $\alpha$.

Fig. \ref{fig:pomfc_vs_na} shows that the learned POMFC-TrueState policy outperforms its finite system version NA policies as well as validating that it outperforms the POMFC policy using average belief instead of true state. 
Further, the figure shows that the POMFC-AvgBelief policy performs better than the NA policy, even when it uses the true AoI. 
Recall, that the NA policies only utilize the AoI of each agent with no additional information. 
This result is an indication the particle filter proposed in the particle filter section works well, as also shown in Figs. \ref{fig:pf_hist} and \ref{fig:particle_filter_perf}. 
Hence, if the true AoI at the receiver is not instantaneously available for learning the policy then making use of the agent beliefs allows to obtain a well-performing policy.

Fig. \ref{fig:na_dec} shows the performance comparison of the three versions of NA-Dec: TrueState, AvgBelief, and Particles. It can be seen that the policies using average belief and particles have a very similar performance and they both outperform the TrueState policy. 
This is because our particle filter is able to maintain an accurate belief over the true state \textit{and} also uses  additional information such as standard deviation over particles or the whole set of particles to learn a more explored and better policy.
\begin{figure}
\centering
\includegraphics[scale=0.49]{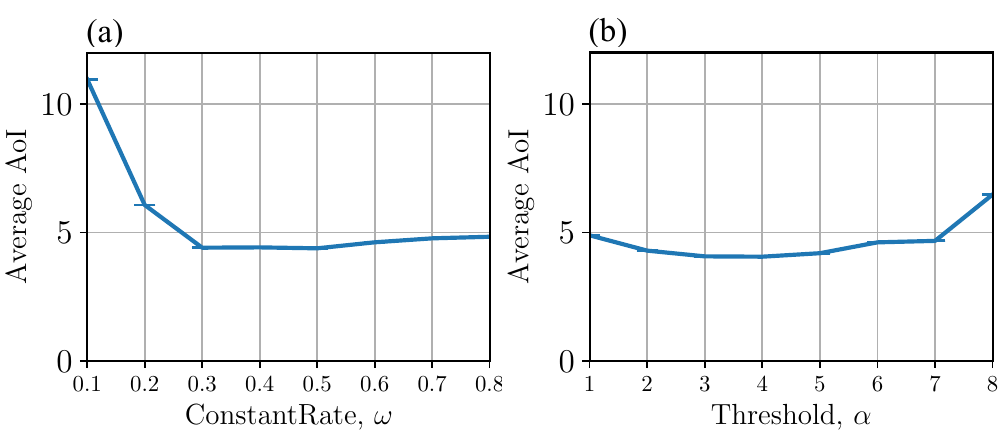}
\caption{(a) The effect of increasing the constant rate $\omega$, i.e., the proportion of uniformly selected agents to send at every time step on the average AoI of the system. (b) The effect of increasing the true AoI threshold $\alpha$ that triggers sending a fresh message. Both plots are for $N=2000$ agents.}
\label{fig:cr_range}
\end{figure}

Fig. \ref{fig:cr_range}(a) shows the average AoI for the ConstantRate policy for a wide range of rates $\omega$. 
The rate $\omega$ denotes the proportion of the agents (chosen uniformly at random) that send at every timestep such that $\omega=1$ is equivalent to the AlwaysSend policy.
It can be seen that there exists a non-trivial optimum over the rate $\omega$. 
We have chosen the constant rate of $\omega=0.5$ for further comparisons to other policies, i.e., at every timestep half of the agents chosen uniformly at random will be sending their messages.
Fig. \ref{fig:cr_range}(b) gives the performance comparison for using different AoI threshold values, $\alpha$, where here too a non-trivial optimum exists.
In further comparisons we use the value $\alpha=4$.
These two results show that a fixed policy of sending at a constant rate or a simple state-based threshold policy cannot yield optimal performance and there is a need for learning a dynamic policy.

\begin{figure}
\centering
\includegraphics[scale=0.49]{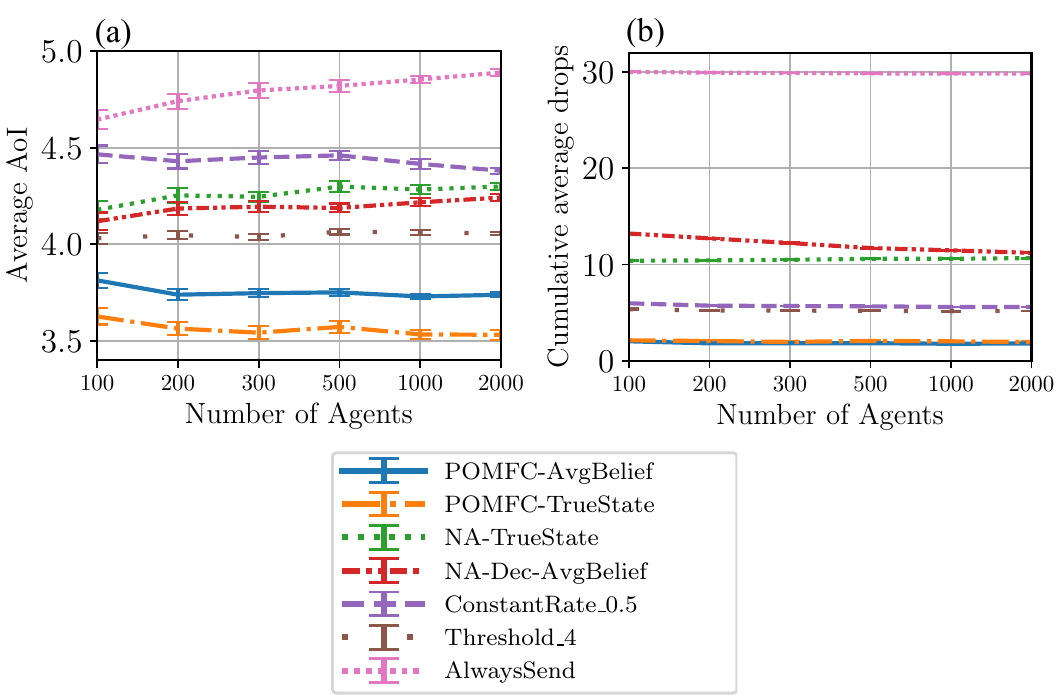}
\caption{{POMFC policies in comparison to the best performing (i) NA policies (NA-TrueState, Na-Dec-AvgBelief), (ii) ConstantRate policy, (iii) Threshold policy, and (iv) AlwaysSend policy. POMFC-TrueState outperforms all others in terms of average AoI as well as the average drops in the system. POMFC-AvgBelief outperforms all except for POMFC-TrueState providing a good policy when the channel delay is unknown but taken into account.
}}
\label{fig:all_comparison}
\end{figure}

Finally, Fig. \ref{fig:all_comparison} shows the performance  of our proposed POMFC policies to the best performing NA (NA-TrueState) and best performing NA-Dec (NA-Dec-AvgBelief). Comparison is also taken to the best performing ConstantRate at $\omega=0.5$ and Threshold at $\alpha=0.4$. It can be seen that our MDP-based POMFC-TrueState policy outperforms all other policies in terms of the average AoI of the system.
Additionally, we show the comparison of the cumulative average number of message drops, where the average is calculated at each timestep. Here also our POMFC-TrueState policy has the least average number of drops, which intuitively goes hand in hand with minimizing the average AoI. 
Finally, our POMFC-AvgBelief policy, which uses our proposed particle filter, performs better than all the other learned and fixed policies (except when knowing the TrueState for POMFC), both in terms of average AoI and average message drops.
Note that AoI is used mostly in time sensitive systems, a few milliseconds can also make a huge difference.
This indicates that our particle filter performs well and is  suitable for training in the setup where true AoI is not instantaneously available to the agents.

\section{Conclusion}
We considered a multi-agent model wherein agents collaboratively minimize the expected Age of Information that is associated with their transmitted messages at the receiver.
The agents simultaneously use a finite capacity random delay-inducing channel to transmit their fresh messages to the receiver, which then uses a similar channel to reply with acknowledgements on the updated AoI.
Since the acknowledgements are received by the agents with a delay the system is treated as partially observable. 
Further, the agents work in a decentralized manner with no communication to coordinate, making them unaware of the global state and actions in the system.
In order to cater to the delayed acknowledgements we proposed a bootstrap particle filter to maintain a belief over the AoI of the agent at the receiver using the received delayed acknowledgements.
Further, in order to solve the decentralized system in a scalable manner, we employ the partially observable mean-field control framework which effectively transforms the decentralized multi-agent system into a single-agent Markov Decision Process. 
We subsequently solve this MDP using the state-of-the-art PPO algorithm. 
Our results demonstrate that our POMFC approach outperforms both learned and fixed policies, showing the efficacy of MFC methods in practical and application-oriented problem settings. 
Our POMFC model using the particle filter also performs better than all other N-Agent learned and fixed policies, making it an interesting direction to explore in the absence of receiver information.

%% file: example_realization_figure.tex
\begin{figure}
\centering
\begin{tikzpicture}[scale=0.9]
\pgfmathsetseed{12}
  \draw[->] (-.5, 0) -- (8, 0) node[below right] {$t$};
  \draw[->] (0, -.5) -- (0, 3.7) node[above left] {$x_{n}(t)$};

 \draw[blue] (0,0) -- (1.5,1.5);
 \draw[blue]  (1.5,1.5) -- (1.5,0.5);

  \draw[blue]  (1.5,.5) node[cross=5pt,red, very thick] {};
  \draw (1.5,.5) node[left] {$z_{n,1}$};
      \draw (1.5, 0.5pt) -- (1.5,-3pt)node[anchor=north] {$\eta_{n,1}$};

 \draw [blue] (1.5,.5) -- (3,2);
 \draw [blue] (3,2) -- (3, 1);

        \draw (3,1) node[cross=5pt,red, very thick] {};
 \draw (3,1) node[left] {$z_{n,2}$};
  \draw (3,1pt) -- (3,-3pt)	node[anchor=north] {$\eta_{n,2}$};

 \draw[blue]  (3, 1)-- (6,3.6);
 \draw[blue] (6,3.6) -- (6,2);
  \draw[blue]  (6,2) -- (7.6,3.6);

    \draw (6,2) node[cross=5pt,red, very thick] {};
 \draw (6,2) node[left] {$z_{n,3}$};
  \draw (6, 1pt) -- (6,-3pt)	node[anchor=north] {$\eta_{n,3}$};

     \draw[fill= green]  (5,0.5) circle (3pt);
  \draw (5.1,0.5) node[right] {$y_{n,1}$};
      \draw (5, 0.5pt) -- (5,-3pt)	node[anchor=north] {$\eta'_{n,1}$};

        \draw[fill= green]  (6.9,2) circle (3pt);
 \draw (6.9,2) node[right] {$y_{n,2}$};
  \draw (6.6, 1pt) -- (6.6,-3pt)	node[anchor=north] {$\eta'_{n,2}$};

    \draw[fill= green] (7.6,1) circle (3pt);
 \draw (7.6,1) node[right] {$y_{n,3}$};
  \draw (7.5, 1pt) -- (7.5,-3pt)	node[anchor=north] {$\eta'_{n,3}$};

  \draw[->, thick] (1.5,.5) -- (4.9,0.5);
  \draw[->, thick] (6,2) -- (6.8,2);
    \draw[->, thick] (3,1) --  (7.5,1) node[pos=0.5, above] {$(\eta_{n,2}'-\eta_{n,2})$};
\end{tikzpicture}
\caption{Realization of the AoI process $x_n(t)$ together with the noise-free measurements $z_{n,j}$.
Observe the 
out-of-order delivery of the acknowledgements to the agent $n$ on the backward channel, i.e., the second ack $y_{n,2}$ arrives at time point $\eta'_{n,2}$ at the sensor $n$ containing the updated AoI $z_{n,3}$  of the third message.}
\label{fig:sample}
\end{figure}

%% file: particle_filter_figure.tex
\begin{figure}[t!]
\centering
\begin{tikzpicture}[scale=0.6]
\pgfmathsetseed{12}
  \draw[->] (-.5, 0) -- (8, 0) node[below right] {$t$};
  \draw[->] (0, -.5) -- (0, 5) node[below left] {$x(t)$};
 \draw[blue] (0,0) -- (1.5,1.5);
 \draw[blue]  (1.5,1.5) -- (1.5,0.5); 
  \draw[blue]  (1.5,.5) node[cross=5pt,red, very thick] {};
  \draw (1.5,.5) node[left] {$z_{1}$};
   \draw (1.5, 0.5pt) -- (1.5,-3pt)node[below right] {$\eta_{1}$};
\draw [blue] (1.5,.5) -- (3,2);
 \draw [blue] (3,2) -- (3, 1);
    \draw (3,1) node[cross=5pt, red, very thick] {};
 \draw (3,1) node[left] {$z_{2}$};
  \draw (3,1pt) -- (3,-3pt)	node[below left] {$\eta_{2}$};
 \draw[blue]  (3, 1)-- (6,3.6);
 \draw[blue] (6,3.6) -- (6,2);
  \draw[blue]  (6,2) -- (7.6,3.6);
    \draw (6,2) node[cross=5pt, red, very thick] {};
 \draw (6,2) node[left] {$z_{3}$};
  \draw (6, 1pt) -- (6,-3pt)	node[below left] {$\eta_{3}$};  
     \draw[fill= green]  (5,0.5) circle (3pt);
  \draw (5.1,0.5) node[right] {$y_{1}$};
      \draw (5, 0.5pt) -- (5,-3pt)	node[anchor=north] {$\eta_{1}'$};  
        \draw[fill= green]  (6.9,2) circle (3pt);
 \draw (6.9,2) node[right] {$y_{2}$};
  \draw (6.9, 1pt) -- (6.9,-3pt)	node[anchor=north] {$\eta'_{2}$};
    \draw[fill= green] (7.6,1) circle (3pt);
 \draw (7.6,1) node[right] {$y_{3}$};
  \draw (7.6, 1pt) -- (7.6,-3pt)	node[anchor=north] {$\eta'_{3}$};
  \draw[->, thick] (1.5,.5) -- (4.9,0.5);
  \draw[->, thick] (6,2) -- (6.8,2);
    \draw[->, thick] (3,1) --  (7.5,1);
    \draw[->] (-0.5, -7) -- (8, -7) node[below right] {$t$};
  \draw[->] (0, -7.5) -- (0, -2) node[below left] {$x_{1}(t)$};
 \draw[orange, thick] (0,-7) -- (1.7,-5.3);
  \draw[orange, thick]  (1.7,-5.3) -- (1.7,-6.2);
  \draw[orange, thick]  (1.7,-6.2) -- (5,-2.9);
  \draw[fill= black] (5,-2.9) circle (2pt);
  \draw[fill= green] (5,-6.5) circle (3pt);
 \draw (5.1,-6.6) node[right] {$y_1$};
 \draw[<-, thick] (1.5,-6.5) -- (4.9,-6.5);
\draw[magenta, dashed]  (1.5,-6.5) -- (3.5,-4.5);
\draw[magenta, dashed]  (3.5,-4.5) -- (3.5,-5.5);
\draw[magenta, dashed]  (3.5,-5.5) -- (5,-4);
\draw[magenta, thick]  (5,-4) -- (6.9,-2.1);
 \draw[fill= black] (6.9,-2.1) circle (2pt);
  \draw[fill= green]  (6.9,-5) circle (3pt);
 \draw (6.9,-5) node[right] {$y_2$};
 \draw[<-, thick] (6,-5) -- (6.8,-5);
  \draw[yellow, dashed]  (6,-5) -- (7, -4);
  \draw[yellow, thick]  (7,-4.) -- (7.6, -3.4);
\draw[fill= green] (7.6,-6) circle (3pt);
 \draw (7.6,-6) node[right] {$y_3$};
  \draw[<-, thick] (3,-6.) -- (7.5,-6);
  \draw[fill= black] (7.6, -3.4) circle (2pt);
  \draw[cyan, dashed] (3,-6.) -- (4., -5);
  \draw[cyan, dashed] (4., -5) -- (4, -5.5);
  \draw[cyan, dashed] (4, -5.5) -- (6.5, -3.);
    \draw[cyan, dashed] (6.5, -3.) -- (6.5, -4);
     \draw[cyan, dashed] (6.5, -4.) -- (7.6, -2.9);
\draw[cyan, thick] (7.6, -2.9) -- (8, -2.5);
    \draw[->] (-0.5, -14) -- (8, -14) node[below right] {$t$};
    \draw[->] (0, -14.5) -- (0, -9) node[below left] {$x_{2}(t)$};
        \draw[olive, thick] (0,-14) -- (2.2,-11.8);
\draw[olive, thick]  (2.2,-11.8) -- (2.2,-13.);
\draw[olive, thick] (2.2,-13) -- (5.,-10.2);
\draw[fill= black] (5,-10.2) circle (2pt);
    \draw[fill= green] (5,-13.6) circle (3pt);
 \draw (5.1,-13.6) node[right] {$y_1$};
  \draw[<-, thick] (1.5,-13.6) -- (4.9,-13.6);
  \draw[cyan, dashed] (1.5, -13.6) -- (5, -10.1);
  \draw[cyan, thick] (5, -10.1) -- (5.5, -9.6);
  \draw[cyan, thick] (5.5, -9.6) -- (5.5, -12.5);
  \draw[cyan, thick] (5.5, -12.5) -- (6.9, -11.1);
  \draw[fill= black] (6.9, -11.1) circle (2pt);
  \draw[fill= green]  (6.9,-12) circle (3pt);
 \draw (6.9,-12) node[right] {$y_2$};
 \draw[<-, thick] (6.,-12) -- (6.8,-12);
 \draw[purple, dashed] (6.,-12) -- (6.9, -11.1);
 \draw[purple, thick] (6.9, -11.1) -- (7.6, -10.4);
 \draw (7.6,-13) node[right] {$y_3$};
\draw[fill= green] (7.6,-13) circle (3pt);
\draw[<-, thick] (3,-13) -- (7.5,-13);
\draw[fill= black] (7.6, -10.4) circle (2pt);
 \draw[orange, dashed] (3.,-13.) -- (5.2, -10.8);
 \draw[orange, dashed] (5.2, -10.8) -- (5.2, -12);
 \draw[orange, dashed] (5.2, -12) -- (7.6, -9.6);
 \draw[orange, thick] (7.6, -9.6) -- (8, -9.2);
\draw[dashed] (0.2, -14.5) -- (0.2, 5) node[above] {$\tau_1$};
\draw[dashed] (1., -14.5) -- (1., 5) node[above] {$\tau_2$};
\draw[dashed] (3.2, -14.5) -- (3.2, 5) node[above right] {$\tau_3$};
\draw[dashed] (1.5, -14.5) -- (1.5, 5) node[above] {$\rho_1$};
\draw[dashed] (6, -14.5) -- (6, 5) node[above] {$\rho_2$};
\draw[dashed] (3, -14.5) -- (3, 5) node[above] {$\rho_3$};
\end{tikzpicture}
\caption{Using the realization from Fig.~\ref{fig:sample}, we visualize two particles, $(x_{1}(t)$, $x_{2}(t)) \in \mathcal X$.
Recall that $\tau_i$ is sending time  of message $i$, $\eta'$ are the reception times of the acks at the receiver, $\rho_i$ are the ordered sequence of acknowledgments,
$x_t$ is the true AoI for this agent at the receiver and $y$ are the received observations at the time points marked by black dots, after which the particles are back-propagated using the received information. From these points on re-simulations, shown using dashed lines, until the next observation time $\eta'$.
}
\label{fig:belief_update}
\end{figure}